\documentclass[journal]{IEEEtran}
\usepackage{fancyhdr,color, graphics, epsfig}
\usepackage{graphicx,subfigure}

\def\figurename{Fig.}

\begin{document}

\title{\emph{UMPIRE}: a Universal Moderator for the Participation in IETF Remote Events}

\author{Simon~Pietro~Romano%
\thanks{S.~P.~Romano is with the Department of Electrical Engineering and Information Technology (DIETI), University of Napoli Federico II, Via Claudio 21, 80125 Napoli, Italy, e-mail: \texttt{spromano@unina.it}% <-this % stops a space}
}}

\maketitle

\thispagestyle{fancy}
\normalfont

\begin{abstract}
UMPIRE provides seamless meeting interaction among remote and local participants. It uses the Binary Floor Control Protocol (BFCP), an IETF standard for moderation. BFCP introduces automated floor control functions to a centralized conferencing environment. The paper discusses the design and the implementation of the UMPIRE system and highlights the most notable solutions we devised to handle variegated requirements and constraints. We also discuss the lessons learned while experiencing in first person how the concrete
application of research results which have eventually brought to new standards still has to confront itself with a number of minor yet concrete issues which might altogether undermine the overall process bringing to wide adoption from the community.

\end{abstract}

\section{Background, rationale and motivation}

UMPIRE provides seamless meeting interaction among remote and local participants. It uses the Binary Floor Control Protocol (BFCP), a standard for moderation. BFCP introduces automated functions for a centralized conferencing environment. The project has been motivated by the ongoing efforts within the IETF (Internet Engineering Task Force), to standardize mechanisms for enabling remote participation at meetings. At that time, work was in full swing and was being formalized in a (now expired) Internet draft\footnote{``Requirements for Remote Participation Services for the IETF'', \texttt{draft-ietf-genarea-rps-reqs-08}} specifically devoted to this task. The draft in question was actually being produced at the request of the IETF Administrative Oversight Committee (IAOC), which issued an ad hoc request for proposals at the end of October $2011$\footnote{http://iaoc.ietf.org/documents/RPS-Specifications-RFP-2011-10-19.pdf}. The draft contains discussion (section 2.3.4.1) dedicated to the task of moderation. Titled \emph{``Floor Control for Chairs for Audio from Remote Attendees''}, it contains a list of requirements, including:
 
\begin{quote}
   **Requirement 08-31**: Remote attendees MUST have an easy and
   standardized way of requesting the attention of the chair when the
   remote attendee wants to speak.  The remote attendee MUST also be
   able to easily cancel an attention request.
   
  **Requirement 08-33**: The floor control portion of the Remote Participation System
   MUST give a remote attendee who is allowed to speak a clear signal when they
   should and should not speak.

   **Requirement 08-34**: The chair MUST be able to see all requests
   from remote attendees to speak at any time during the entire meeting
   (not just during presentations) in the floor control system.
   
   **Requirement 08-35**: The floor control system MUST allow a chair to
   easily mute all remote attendees.

   **Requirement 08-36**: The floor control system MUST allow a chair to
   easily allow all remote attendees to speak without requesting
   permission; that is, the chair SHOULD be able to easily turn on all
   remote attendees mics at once.  
\end{quote}   
   
UMPIRE is currently capable of meeting most of the requirements in that list.

We also took inspiration from related mailing list discussions. To get an idea of such discussions, consider the following thread on the IETFÕs Working Group Chairs mailing list: \textit{``Getting Taipei remote participants' input''}. It includes a message from the author of this paper\footnote{\small{http://www.ietf.org/mail-archive/web/wgchairs/current/msg10274.html}}, who was proposing to use RFID (Radio Frequency IDentification) tags to trigger BFCP requests by in-person participants, in much the same way as the floor requests generated by remote participants. In answer to such message, Dave Crocker wisely stated the following\footnote{\small{http://www.ietf.org/mail-archive/web/wgchairs/current/msg10277.html}}:

\begin{quote}
Suggestions like the above sound appealing. Unfortunately, they are far beyond current products and making them useful is considerably more difficult than the suggestions imply. That doesn't mean they should be ignored, but we need to be careful about slipping into the assumption that merely citing a bit of technology means that an issue is resolved. We had an experiment with RFIDs. It was awkward, at best. In the case of queue management, we have at least entering the queue, position in the queue, and the chair's control of the queue.
\end{quote}

A further fundamental contribution to the discussion was also provided by Brian Rosen, who asserted\footnote{\small{http://www.ietf.org/mail-archive/web/wgchairs/current/msg10280.html}}:

\begin{quote}
People worry about RFID, but I like it because it's a faster read. All I think you want is a reader and a visible queue.
The queue just tells you that the reader read correctly and has you in the queue. The chair gets to change the queue, but that ought to be rare and probably just pick the next person in queue. Remote participants simply imitate the reader action.
\end{quote}

The tool could also provide session chairs with the ability to grant ``business class'' requests (i.e., in the case of cut and thrust debates, or in the presence of intervention of an Area Director) so that individuals obtain higher priority, essentially putting such requests on top of the queue. This highlights the possibility of applying different activity templates, or paradigms, for common handling of remote and in-person participants, according to different group process modes. 

We took into account these considerations and applied the usual IETF approach of having a running code prototype to identify and clarify possible issues and foster discussion. The system we developed provides a proof of concept for a moderation framework built on top of the Meetecho conferencing system~\cite{Meetecho}. Meetecho is a standards-based conferencing architecture used at IETF meetings for remote participation.

The remainder of the paper is hence organized as follows. Section~\ref{sect:Context} helps the reader position this work in the context of ongoing IETF standardization efforts for multimedia conferencing, while Section~\ref{sect:Related} provides some information about related work in the field of conference moderation. Section~\ref{sect:Architecture} describes the overall architectural design of the UMPIRE system, whose implementation is briefly sketched in section~\ref{sect:Implementation}. Section~\ref{sect:LessonsLearnt} discusses the main issues we faced during the design and implementation phases and provides insight in the lessons learnt. Finally, Section~\ref{sect:Conclusions} summarizes the discussion stimulated by UMPIRE within the IETF community, illustrating the useful feedback we gathered and the envisaged direction of our future efforts.

\section{Context}
\label{sect:Context}

The IETF has devoted many efforts to the specification of standard
conferencing solutions. They include the Framework for Centralized Conferencing (XCON Framework)~\cite{XCON-Framewrok},
which defines a signaling-agnostic architecture, naming conventions and logical entities required for building advanced conferencing
systems. An XCON-compliant framework architecture comprises several protocols, including the Binary Floor Control Protocol (BFCP)~\cite{BFCP} which is associated with all moderation operations for a conferencing session.

\begin{figure}
    \centering
        \includegraphics[width=0.90\columnwidth]{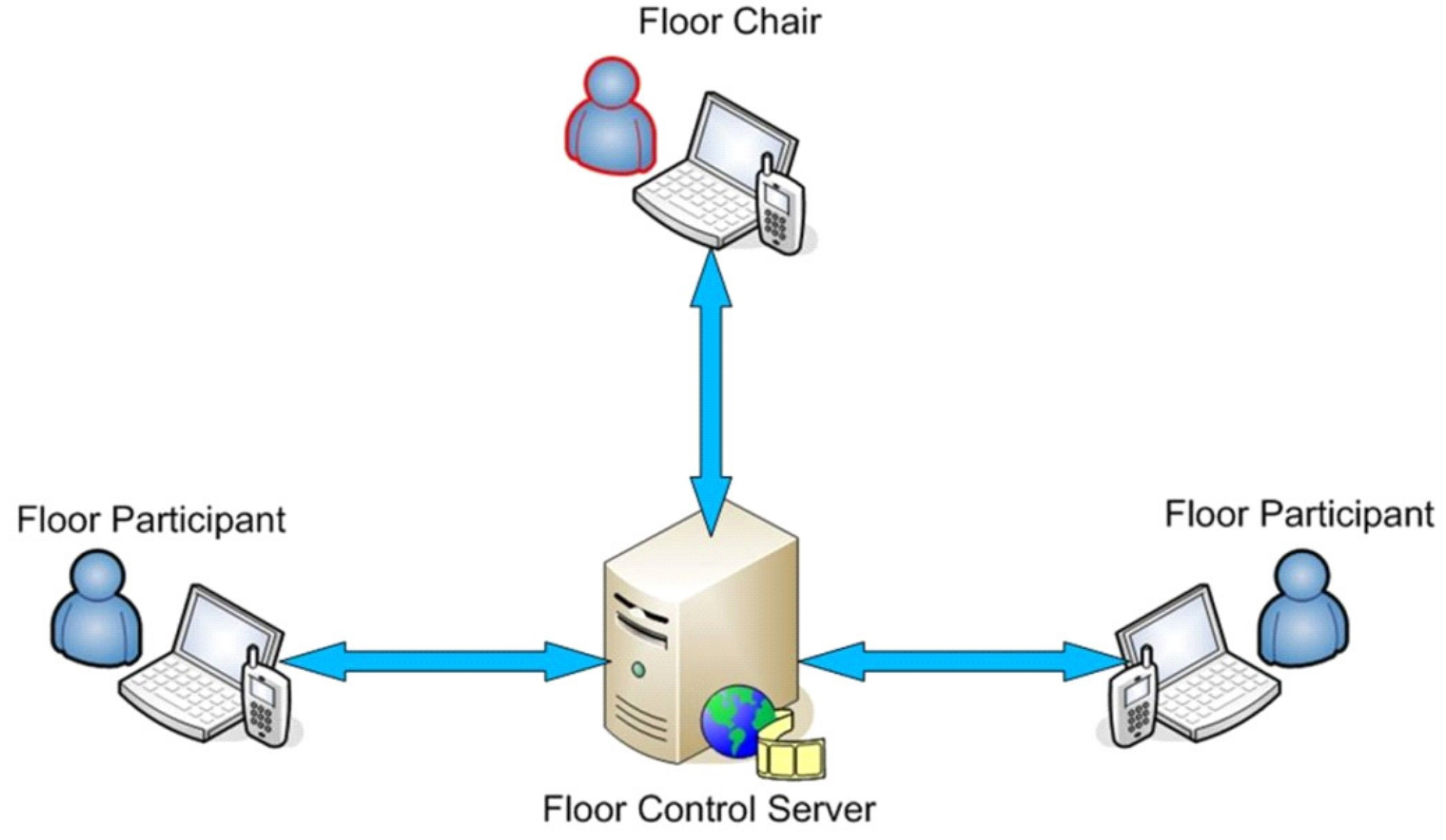}
    \caption{The Binary Floor Control Protocol architecture}
    \label{fig:BFCP_0}
\end{figure}

As depicted in \figurename~\ref{fig:BFCP_0}, BFCP models the presence of floor participants who ask for access to the conference floor (e.g., audio and/or video) by sending messages to a central entity called floor control server. The server itself does not make decisions on its own, but rather forwards requests to the floor chair, who acts as a moderator and is in charge of taking decisions.

When a conference participant asks for the floor, it sends a request message to the floor control server, which forwards it to the floor chair. When the chair makes a decision, it informs the floor control server, which in turn notifies both the requesting participant and all other participants potentially interested in receiving floor control notifications.

\section{Related Work}
\label{sect:Related}

Floor control has since long been the subject of a number of researches. Before BFCP saw the light, some interesting approaches have been proposed in the literature. The authors in~\cite{Aceves2005} propose to use floor control in videoconferencing applications as a means to improve their scalability. With the proposed approach, moderation allows to keep control over both processing and communication overhead, by allowing a maximum of two simultaneous streams at a time and hence mimicking a two-party videoconference. A different approach is embraced in~\cite{Banik2005}. Here, the authors propose an implementation of ALOHA and DQDB (Distributed Queue Dual Bus), two well-known MAC access protocols for local area networks, in a distributed overlay setting. 
The authors of~\cite{Jzer2008} instead design an architecture for medium-sized peer-to-peer conferences. The system is equipped with a floor control mechanism to prevent too many users from speaking simultaneously (hence degrading the audio quality). To the purpose, a distributed role-based floor control protocol is introduced. The protocol leverages floor utilization statistics in order to optimize floor management activities.

More recently, the work in~\cite{Rubaye2009} has proposed an effective way to bring the functions made available by the BFCP protocol to the IP Multimedia Subsystem standard framework. They actually propose to use BFCP for the implementation of the IMS \texttt{Fc} interface envisaged by the $3$GPP standard.

\section{UMPIRE Architecture Design}
\label{sect:Architecture}

UMPIRE fills the role of floor chair in a conference. More precisely, when a conference starts, the UMPIRE user will log in as floor chair.
Subsequent floor requests from conference participants will be:

\begin{enumerate}
  \item stored (in a PENDING state) in the First-Come-First-Served queue at the centralized floor control server;
  \item forwarded to UMPIRE, which will make decisions by assigning the floor to one or more users, thereby triggering state changes at the server. As an example, if the floor control policy has been configured to grant the floor to one user at a time and the UMPIRE accepts, in sequence, three requests coming from three different users, the following will happen:  
  
      \begin{itemize}
        \item the first accepted request will move from a PENDING state to a GRANTED state;
        \item the second and third requests will move from a PENDING state to an ACCEPTED state, indicating that they are ready to be served as soon as the currently GRANTED request has terminated (through either a floor release action from the client or a revoke action from the UMPIRE himself);
        \item when the GRANTED request is completed, the first ACCEPTED request in the queue is granted the floor, while the second one  becomes the ready-to-be-granted request in the floor control server queue;
        \item if the server policy allows for a maximum of $n$ requests to be granted the floor at the same time, up to $n$ clients will reach the GRANTED state and will share the floor in question (e.g., in an audio conference, they will be all allowed to talk at the same time by contributing to the audio mix produced at the conference server).
      \end{itemize}
  \item acknowledged to the clients through BFCP notifications (which allow participants to be kept up to date with respect to the state of their floor transactions).
\end{enumerate}

The above scenario relies on the presence of a centralized floor control server with a queue managed by the chair and containing all floor requests coming from conference participants. This simple model allows for the introduction of advanced moderation functionality.

The interesting thing about the model is that the BFCP queue can be populated with requests coming both from remote participants (equipped with BFCP-enabled clients) and local participants, thanks to the utilization of an agreed-upon procedure for requesting access to the conference floors when one is physically present in the conference room. The usual way of gaining the audio floor at regular IETF meetings is for local participants to politely wait (in a First-Come-First-Served queue) at one of the conference room microphones, in order to either ask for questions or provide their own view on the topic being discussed. If the microphones themselves were equipped with some simple means for:

\begin{enumerate}
  \item recording the presence of users;
  \item sending a trigger (i.e., a floor request) to the floor control server  every time a new user lines up at the microphone
\end{enumerate}

The floor control server queue could transparently (and democratically) moderate a conference envisaging the contemporary presence of both local and remote participants. Indeed, this is what we implemented. With respect to the means for recognizing the presence of users lining up at the microphone, we decided to rely on the RFID technology. We opted for the following policy: (i) an RFID reader was placed close to each of the conference room microphones; (ii) an RFID tag was assigned to each local participant willing to actively participate in the mechanism.

With this approach, when a local participant wanted to contribute to the ongoing discussion, all they had to do was to let their RFID tag be read by the RFID reader associated with the microphone at which they lined up. As soon as the participantÕs tag was read, the reader would send a BFCP floor request to the conference chair and let the participant be inserted in the centralized BFCP queue.

An image of the BFCP queue, updated in real-time, was always projected on a screen available in the room, for inspection by meeting participants. The chair of the conference (i.e., the UMPIRE user) was allowed to manage the BFCP queue, by making decisions to grant the floor to individuals. This demonstrated straightforward moderation functionality, for a conference involving both remote and local participants.

\section{UMPIRE implementation}
\label{sect:Implementation}

UMPIRE has been implemented as a Web$2.0$ application, i.e., a dynamic, highly interactive, web-based system. It is based on a bidirectional HTTP communication channel between the UMPIRE participant, acting as the floor chair in a conference, and the floor control server. The channel uses the COMET Server Push approach, as made available by the ZK framework\footnote{\small{http://www.zkoss.org/}}.
Notifications from the server asynchronously arrive at the UMPIRE client and are represented on a web page providing an always up-to-date snapshot of the BFCP queue (with client requests and related BFCP states). Proactive actions undertaken by the UMPIRE (e.g., accepting or denying a PENDING request, or revoking a floor currently GRANTED to one of the participants), are immediately communicated to the floor control server, affecting the BFCP queue, as well as trigger floor notifications which are sent to clients.

\subsection{Sample call flow}

\begin{figure*}
    \centering
        \includegraphics[width=0.7\textwidth]{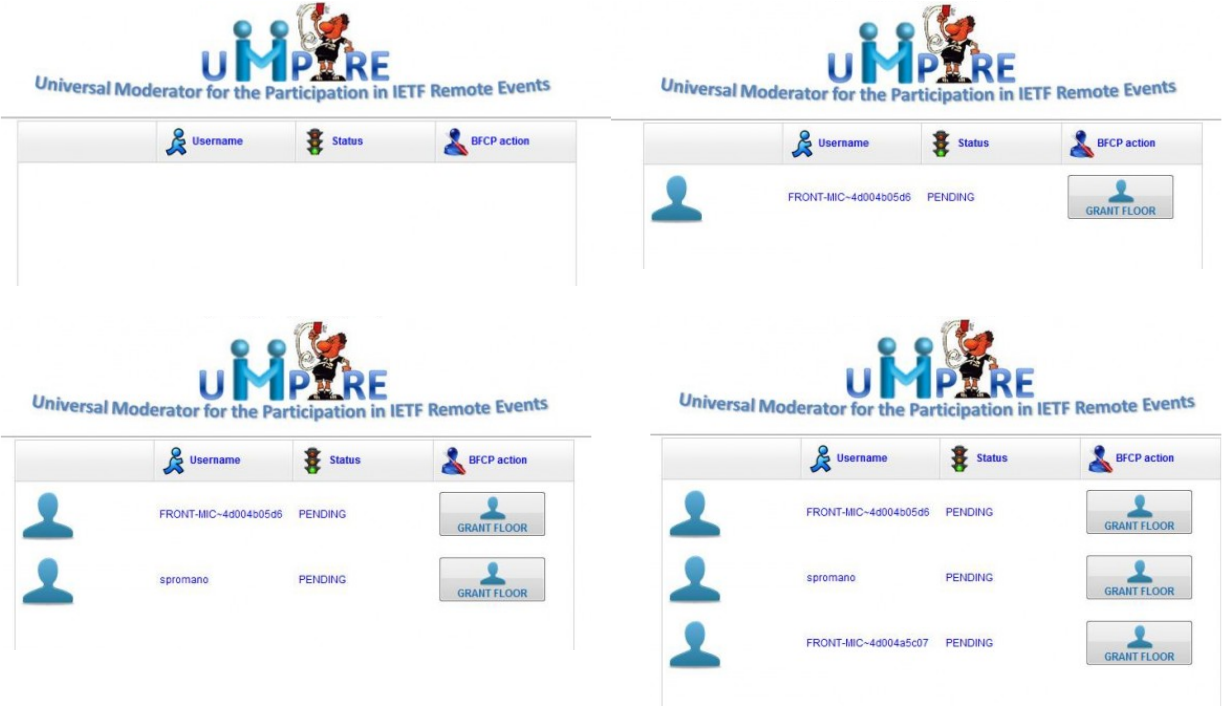}
    \caption{Two local participants and a remote participant asking for the floor}
    \label{fig:CallFlow}
\end{figure*}

The UMPIRE functionality is demonstrated by the following simple scenario:

\begin{itemize}
  \item Three users participate in a conference room:
      \begin{itemize}
        \item two local users, equipped with RFID tags:
            \begin{itemize}
              \item User$1$: $4d004b05d6$
              \item User$2$: $4d004a5c07$
            \end{itemize}
        \item a remote user (whose nickname is `spromano'), who enters the conference through a BFCP-enabled client.
      \end{itemize}
  \item The order in which the three users ask for the audio floor is the following:
        \begin{itemize}
          \item  User$1$ $\rightarrow$ spromano $\rightarrow$ User$2$
        \end{itemize}
\end{itemize}

The situation described above is illustrated in \figurename~\ref{fig:CallFlow}, which shows, respectively, the initially empty queue (top left), the first request arriving from `User1' (top right), the request from the remote participant `spromano' (bottom left), and the final request from local user `User2' (bottom right). The final state of the BFCP queue, after all these actions have been performed, shows the three users (in order of arrival) in a PENDING state, i.e., waiting for the chair to take actions.

\begin{figure*}
    \centering
        \includegraphics[width=0.65\textwidth]{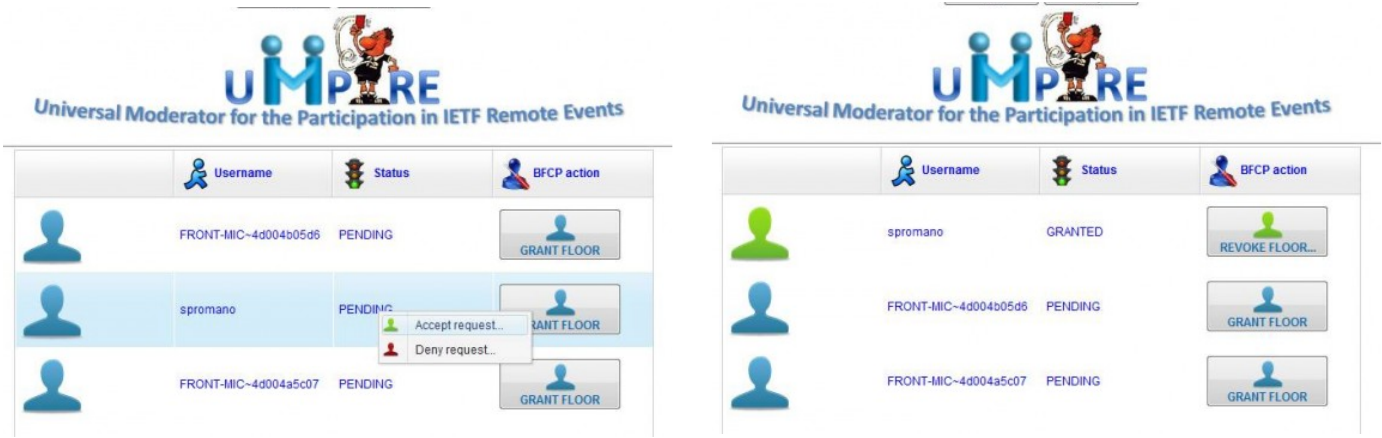}
    \caption{Accepting floor request coming from user `spromano'}
    \label{fig:CallFlow_1}
\end{figure*}

UMPIRE first decides to grant the floor to `spromano', as shown in the two snapshots in  \figurename~\ref{fig:CallFlow_1} (`Accept', left snapshot), which translates into the following actions:

\begin{enumerate}
  \item The BFCP queue is modified: `spromano' becomes first;
  \item The state of the BFCP queue is modified: the audio floor is GRANTED to `spromano' (right snapshot);
  \item remote user `spromano' is unmuted.
\end{enumerate}
    
UMPIRE now decides to grant the floor also to `User2'.

\begin{figure*}
    \centering
        \includegraphics[width=0.75\textwidth]{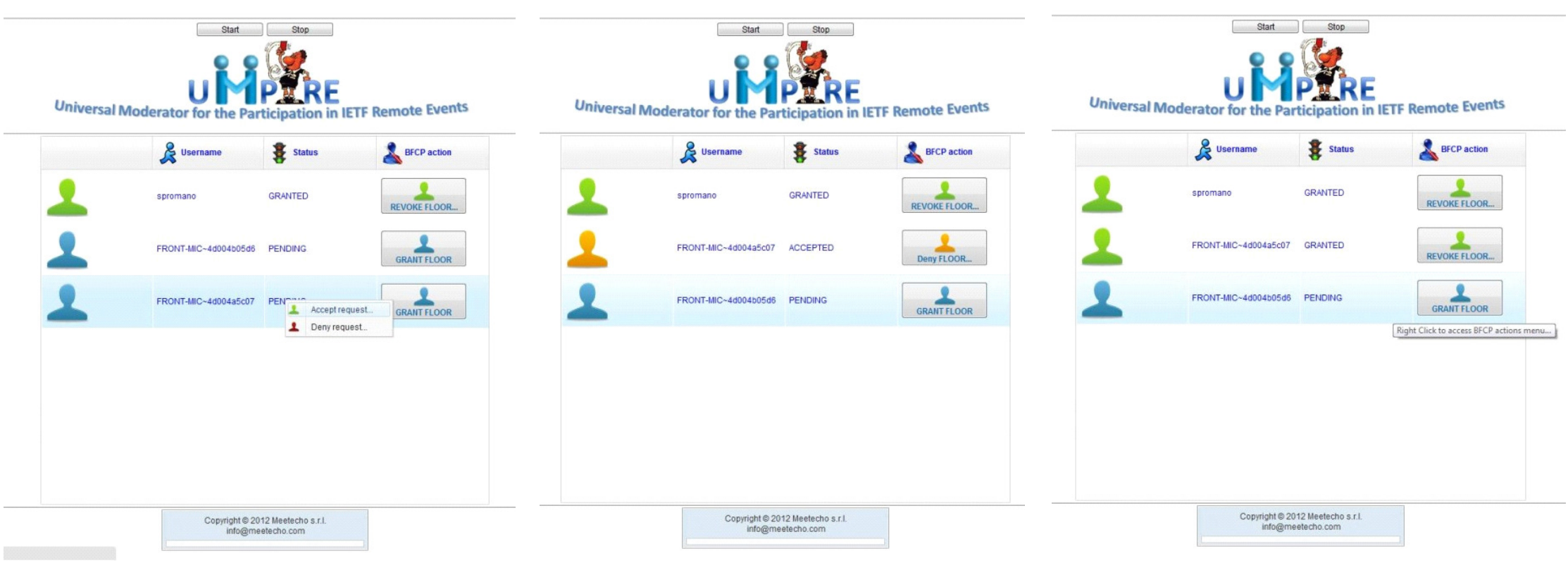}
    \caption{Accepting the request coming from the 'second' local participant}
    \label{fig:CallFlow_2}
\end{figure*}

This is illustrated in \figurename~\ref{fig:CallFlow_2}. 

We can observe from the pictures that the following things have happened:

\begin{enumerate}
  \item UMPIRE has accepted the request issued by `User2' (left snapshot);
  \item The state of the BFCP queue changes: `User2' passes from PENDING to ACCEPTED (middle snapshot) and eventually to GRANTED (right snapshot). This lets us understand that the conference in question has been configured to allow multiple users to be granted the audio floor at the same time. Were this not the case, `User2' would have moved from PENDING to ACCEPTED and would have stayed in such a state as long as the floor was held by `spromano'.
\end{enumerate}

UMPIRE now decides to revoke the floor previously assigned to `spromano'.
    
\begin{figure*}
    \centering
        \includegraphics[width=0.75\textwidth]{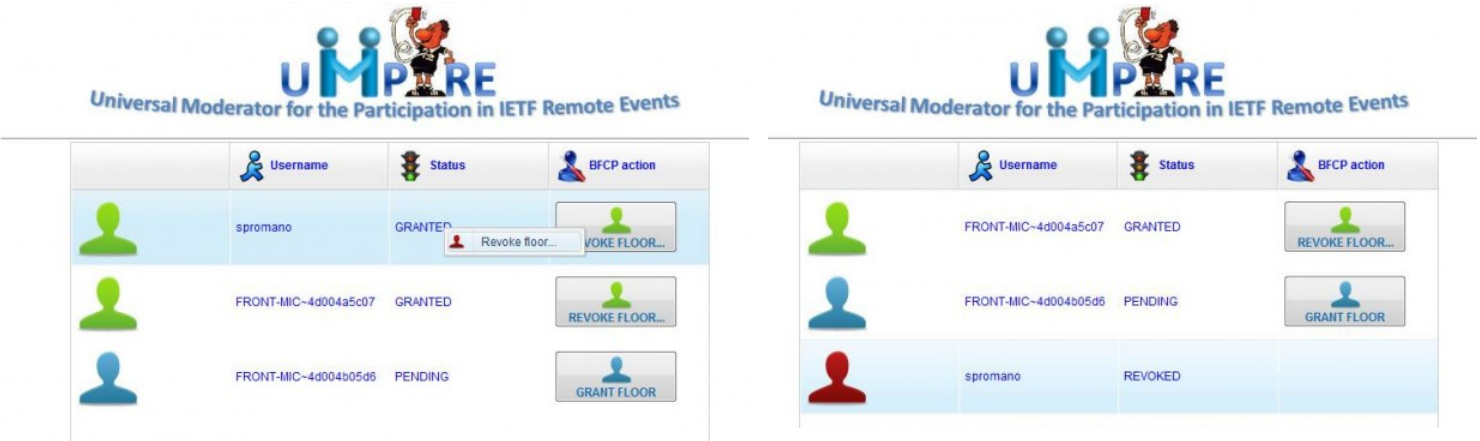}
    \caption{Revoking the floor to user `spromano'}
    \label{fig:CallFlow_3}
\end{figure*}    

This is shown in the snaphots in \figurename~\ref{fig:CallFlow_3}, associated, respectively, with the action undertaken by the chair (left frame) and the effect it entails on both the BFCP queue at the server and the web interface (right frame), which now reports `spromano' in red, with the related REVOKED status.

\section{Lessons Learnt}
\label{sect:LessonsLearnt}

UMPIRE can be regarded as an advanced chapter of the author's experiences with complex conferencing environments, such as  the IETF, dating back to $2005$. This chapter represents a clear example of the way research activities can be brought to the real world, if a proper engineering approach is embraced. The work done to develop the moderation platform taught us several lessons.

From the design perspective, our experience demonstrates the benefits in using a Ôseparation of concernsÕ pattern for re-using a specific subset of the functionality in a large and complex system. Originally we were conducting research on the scalability of conferencing frameworks and hence seized the opportunity to contribute to the ongoing standardization work within the IETF, by actively participating in both the Centralized Conferencing (XCON) and the Media Control (mediactrl) working groups.

The eventual result was Meetecho, a spin-off of the University of Napoli offering a standards-based, conferencing system used to support remote participation in IETF meetings. UMPIRE is built upon Meetecho's major component dealing with BFCP moderation, which was implemented by our research group as a joint activity with Ericsson Nomadic Lab in Helsinki. Notably, it has been conceived as an independent component, which can interoperate with any other BFCP-compliant system; Meetecho is not required. Hence we could focus solely on the client-side of the architecture. Besides re-using the code, we were also able to take advantage of the moderation serverÕs performance, which we have studied in detail in our previous work~\cite{Meetecho}.

As the BFCP chair UMPIRE is in charge of managing requests arriving from conference participants, while keeping an up-to-date representation of the queue, for the web-enabled GUI. The GUI itself was the most challenging part of the overall prototype, since it represents a typical example of a bidirectional, HTTP-based component and must be capable of managing input events coming both from the web interface (e.g., when the moderator clicks on a userÕs icon in order to perform a specific moderation action) and the server-side counterpart residing on the Meetecho conferencing server. For this part, we initially decided to use long polling HTTP requests sent by the client and responsible for the asynchronous update of the web view. This solution proved to be far from optimal, due to the unavoidable overhead for this approach. We then moved to a Comet server push approach, as described earlier. This is definitely better than polling, when an application needs low latency events delivered from the server to the browser. Instead of repeatedly polling for new events, Ajax applications with Comet rely on a persistent HTTP connection between server and client.

Also worth mentioning is the communication between the RFID readers (that we install close to the room microphones) and the BFCP server. This part of the system is critical, since it raises both hardware and software issues. The reader has to interface with passive RFID tags (such as those `embedded' into conference badges worn by participants). It also has to properly communicate with the moderation server, by acting as a standard BFCP participant. Ultimately we used an effective, low-cost, programmable RFID reader made available by Phidgets Inc.\footnote{www.phidgets.com}, a Canadian company offering solutions for the rapid prototyping of RFID sensing components. Fortunately they come with built-in support for an application programming interface written in a number of different languages and freeing developers from all low-level issues associated with RFID sensing capabilities. So the programmer can focus on application-level issues. In our case, we leveraged the Java API.

\begin{figure*}
    \centering
        \includegraphics[width=0.75\textwidth]{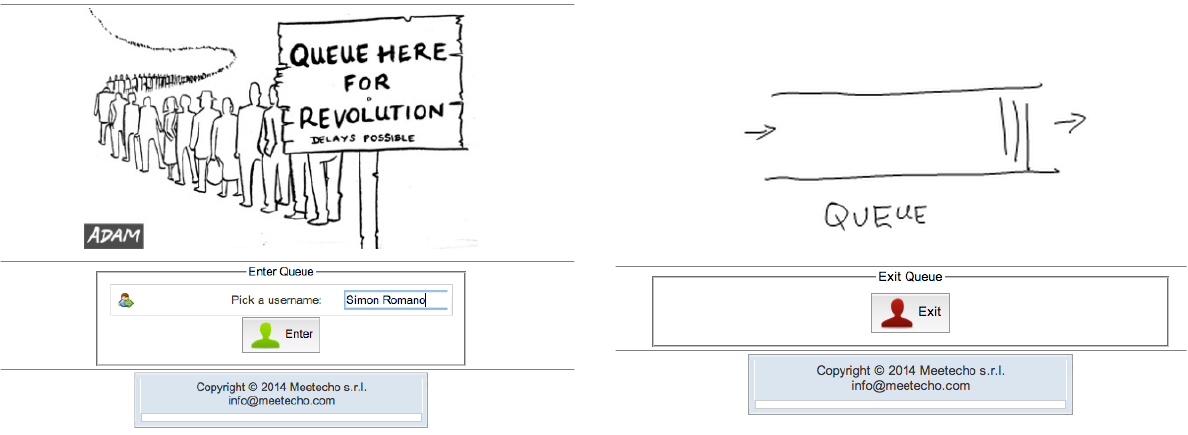}
    \caption{The simple web GUI for sending floor request and floor release messages}
    \label{fig:Web_GUI}
\end{figure*}

As a final remark, we also notice that the idea of enabling RFID-based conference moderation keeps on representing a critical aspect of the UMPIRE system. As part of the feedback that we received after the first experiments at recent IETF meetings, we realized that many people would actually prefer a more lightweight approach to the interaction of conference participants (both local and remote) with the floor control server. Other participant signaling mechanisms are indeed possible, such as through personal smartphones. This suggested us additional experiments to consider, with the final goal to allow the possibility that an operational service would permit multiple means for participants to signal their desire for the floor. Based on the previous consideration, we recently made a further development step and implemented a simple cross-device (laptop, mobile, tablet, etc.) web-based floor control client which can be leveraged by conference participants to send `RFID-less' floor request and floor release messages to the floor control server. With this extension, the UMPIRE system can now moderate a unified `virtual' queue grouping together local and remote participants who make use of either the RFID mechanism, or a BFCP-enabled conferencing application like Meetecho, or the simple Web GUI showed in \figurename~\ref{fig:Web_GUI}. The enhanced version of the UMPIRE system has been used at the $90^{th}$ IETF meeting in Toronto (July $2014$).

\section{Discussion and concluding remarks}
\label{sect:Conclusions}

We have presented UMPIRE, a system for the automated management of floor control and moderation in a meeting room supporting the contemporary presence of local and remote participants. At the time of this writing, UMPIRE (which was first proposed at the $83^{rd}$ IETF meeting in Paris, in March $2012$) has not yet been used `in the wild' as a system to moderate actual meeting sessions. Though, at the last meetings it has been demo-showed during the official ``Meetecho tutorial for Participants and WG Chairs'' and has gathered consensus and appreciation, besides stimulating useful feedback from the audience.

We are currently fine-tuning UMPIRE's functionality to make it ready for official adoption within the IETF. Obviously, nothing prevents UMPIRE from being used in contexts other than the IETF. It can in fact be employed wherever a plea for moderated access to shared resources exists.

\begin{biography}[
{\includegraphics[width=1in,height=1.25in,clip,keepaspectratio]{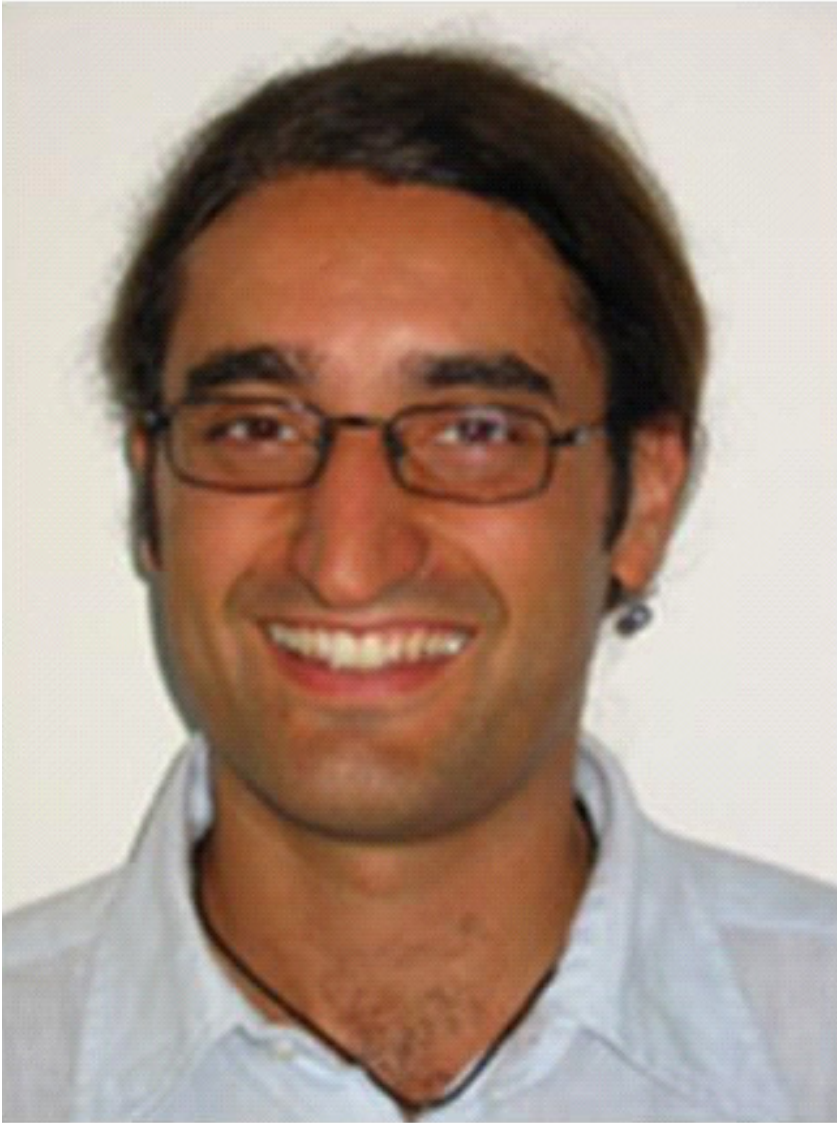}}]
{Simon Pietro Romano} is an Associate Professor in the Department of Electrical Engineering and Information Technology (DIETI) at the University of Napoli. He teaches Computer Networks and Telematics Applications. He is also the cofounder of Meetecho, a startup and University spin-off dealing with WebRTC-based unified collaboration. He actively participates in IETF standardization activities, mainly in the Real-time Applications and Infrastructure (RAI) area.
\end{biography}

%\bibliographystyle{abbrv}
%\bibliography{umpire}

%\balancecolumns

% That's all folks!
\end{document}